\documentclass[twocolumn,showpacs,preprintnumbers,amsmath,amssymb]{revtex4}
\usepackage{graphicx}
\usepackage{dcolumn}
\usepackage{bm}
\usepackage{subfigure}

\begin{document}

\title{Naming Game on small-world networks: the role of clustering structure}
\author{Bo-Yu Lin$^1$}
\author{Jie Ren$^1$}
\author{Hui-Jie Yang$^1$}
\author{Bing-Hong Wang$^{1,2}$}
\email{bhwang@ustc.edu.cn}
\affiliation{%
$^{1}$Department of Modern Physics and Nonlinear Science Center,
University of Science and Technology of China, Hefei 230026,
China\\
$^{2}$Shanghai Academy of System Science, Shanghai 200093, China
}%

\begin{abstract}
Naming Game is a recently proposed model for describing how a
multi-agent system can converge towards a consensus state in a
self-organized way. In this paper, we investigate this model on the
so-called homogeneous small-world networks and focus on the
influence of the triangular topology on the dynamics. Of all the
topological quantities, the clustering coefficient is found to play
a significant role in the dynamics of the Naming Game. On the one
hand, it affects the maximum memory of each agent; on the other
hand, it inhibits the growing of clusters in which agents share a
common word, \emph{i.e.}, a larger clustering coefficient will cause
a slower convergence of the system. We also find a quantitative
relationship between clustering coefficient and the maximum memory.
\end{abstract}


\maketitle

\section{INTRODUCTION}

The semiotic dynamics is a new emerging field concerning the
origination and evolution of languages \cite{L.Steels1997, S.Kirby}.
This field studies how semiotic relations originate and how they
spread in a group of individuals. One of the most interesting
questions in this area is whether and how the group of agents can
converge towards a final consensus state despite an initial
dissension. The Naming Game, as a minimal model, was proposed
recently to study such convergence process \cite{L.Steels1998}. The
original model of Naming Game was presented by Steels \emph{et al.}
in a well-known artificial intelligence experiment called Talking
Head \cite{L.Steels1995}. In the primary experiment, humanoid robots
observed the ambient objects via their digital cameras, and assigned
them different names. It has been certified by the experiment that,
under certain rules, the system will evolve to a consensus state,
characterized by a shared lexicon among robots, without any human
intervention, \emph{i.e.}, the system can converge towards a state
of consensus in a self-organized way.

In this paper, we focus on a special model of Naming Game proposed
by Baronchelli \emph{et al.} \cite{Sharp}. This model characterizes
a non-equilibrium dynamical process of a complex adaptive system and
allows the system to converge towards a special attractor in a
self-organized way, so that, it has gained special attentions from
physicists. This model has been previously investigated on some
typical networks, such as fully connected graphs \cite{Sharp,
Strategy}, regular lattices \cite{Lattice}, small-world networks
\cite{Smallworld}, and heterogenous networks \cite{Non}. These works
uncovered many interesting properties of Naming Game. Recent studies
on network science suggest that network-based dynamics depend
significantly on some special topological properties \cite{Barabasi,
Newman, Boccaletti}. In the context of Naming Game, it has been
found that the topological features of underlying network, such as
the degree distribution, the clustering coefficient, the modularity
and so on, play important roles in the dynamics \cite{Non}. In this
paper, we focus on the clustering structures of the network. We
study the model mentioned above on homogeneous small-world networks
and investigate the relationship between clustering coefficient and
some dynamical quantities of Naming Game.

This paper is organized as follows. In section 2, we describe the
model mentioned above. In section 3 we provide both the simulation
and analytical results. And finally, in section 4, we sum up this
article and discuss some interesting issues that may be relative to
the further works.

\section{MODEL}
The model proposed by Baronchelli \emph{et al.} \cite{Sharp}
captures the essential properties of current dynamics: the system
converges towards a consensus state in a self-organized way. To keep
the paper self-contained, we review the model briefly. The model
includes a set of agents occupying the nodes of a network. Two
agents are permitted to negotiate if they are connected by a link.
For simplicity, it is assumed that there is only one object in the
environment.

At each time step, the present model evolves through the following
rules:

(1)An agent is randomly selected from the network to play the role
of ``speaker", and from the speaker's neighbors, another agent is
randomly selected to be a ``hearer";

(2)The speaker picks up a word, if there is any, from its local
vocabulary and sends it to the hearer; if the speaker's vocabulary
is empty, the speaker will invent a new word to name the object, add
the new word to its own vocabulary, and then send it to the hearer;

(3)After receiving the word from the speaker, the hearer searches
its local vocabulary to check whether it has already been recorded
previously. If this is true, the negotiation will be deemed as a
successful one, and both hearer and speaker will then delete all
other words in their vocabularies; otherwise, the hearer will add
the new received word to its local vocabulary without deleting any
other words.

This process continues until there is only one word left. It is
obvious that at the beginning, the system contains a variety of
different names, however, it has been certified that, after enough
steps of local interactions, the system will eventually share an
exclusive word.

Networks adopted here are the so-called homogeneous small-world
networks(HoSW) proposed by Santos \emph{et al.} \cite{HoSW}.
Slightly different from the original process, we construct our
networks in the following way. Starting from a ring of $N$ nodes,
each connected to its $K$ nearest neighbors, we choose a node and
the edge that connects it to its nearest neighbor in a clockwise
sense. With probability $p$, we exchange this edge with another
unchanged edge (Detailed rules can be seen in Ref. \cite{S.Maslov});
otherwise, we leave it in place. We repeat this exchange by moving
clockwise around the ring until one lap is completed. Afterward, we
consider the edges that connect nodes to their second-nearest
neighbors. As before, we exchanged each of these edges with an
unchanged edge with probability $p$. We continue this process,
circling around the ring and proceeding outward to further neighbors
after each lap, until each edge has been visited once. We forbid any
self-edges and multi-edges during this reconstruction.

Since the degree of each node keeps unchanged during the
reconstruction process \cite{S.Maslov, HoSW}, every node in HoSW
network is of degree $K$. In our simulations, we keep $K=8$ fixed.
Similar to the Wattts-Strogatz(WS) networks \cite{WS, HoSW}, when
the exchange probability $p$ is close to 0, a HoSW network has a
clustering coefficient close to a regular lattice's, and when $p$
increase to the order of $10^{-1}$, a dramatic decrease of the
clustering coefficient can be observed. However, it is still worth
while to mention that network topologies in different backgrounds
may differ from each other, and the reason we adopt HoSW networks is
that, as mentioned above, they are able to bring an effective
quantitative change of clustering coefficient together with
preserving the degree of each node.

\section{NAMING GAME ON HoSW NETWORKS}
In this section, based on the model mentioned above, we investigate
how dynamical properties of Naming Game response to the change of
the network topology. Former works \cite{Non} on Naming Game had
revealed that clustering coefficient and node degree are two of the
most important factors for the dynamics. In this paper, we perform
Naming Game on HoSW networks to eliminate the influence of varying
degrees, and lay our special emphasis on the clustering coefficient.
The local clustering coefficient of node $i$ is defined as $C_{i} =
2e_{i}/k_{i}(k_{i}-1)$, where $e_{i}$ is the number of edges among
$i$'s neighbors, and $k_{i}$ is the degree of node $i$ (in this
paper, $k_{i} \equiv K, \forall i$), and the average clustering
coefficient of the network is $C = \frac{1}{N} \sum_{i} C_{i}$,
where $N$ is the number of nodes, and $i$ runs over all the nodes
\cite{WS}. Figure 1 shows the average clustering coefficient of
networks with different sizes as a function of rewiring probability
$p$. It indicates that the clustering coefficient is almost
independent of the network size.
\begin{figure}
    \centerline{\includegraphics*[width=0.5\textwidth]{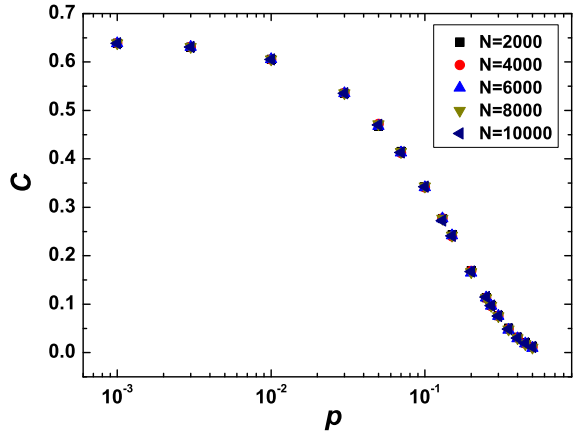}}
    \caption{(Color Online) Clustering coefficient as a function of
    rewiring probability $p$. For each $N$ and $p$, the clustering coefficient is averaged over 20
    independently generated networks.
    }\label{c(p).fig}
\end{figure}

Before discussing the influence of topological quantities on the
dynamical properties of Naming Game, a general view of the dynamics
may be helpful. Two of the most relevant quantities for describing
the dynamical process are the total number of words over the network
$n(t)$, and the number of distinct words over the network $nd(t)$.
\begin{figure}
    \centerline{\includegraphics*[width=0.5\textwidth]{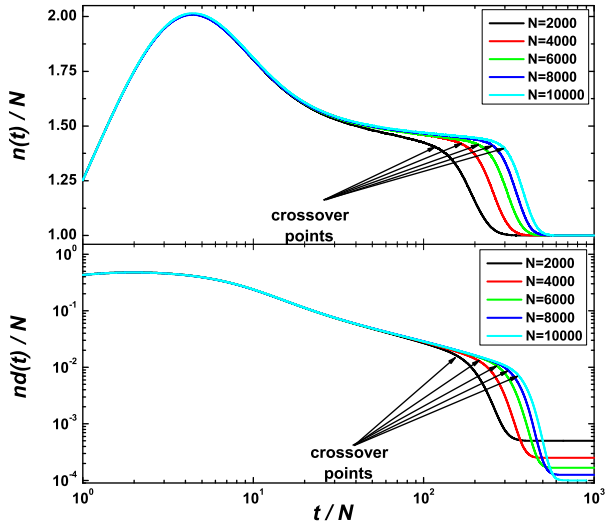}}
    \caption{(Color Online) The average words per agent $n(t)/N$ as a function of
    rescaled time $t/N$. $p=0.1$, averaged over 5000 realizations realizations. The
    maximum value of $n(t)$ is proportional to $N$. (Color Online) The evolution of the rescaled number of different words $nd(t)/N$ for
    different network sizes a function of time. $p=0.1$, averaged over 5000 independent
    realizations.
    }\label{n(t).fig}
\end{figure}

Figure 2 displays the evolution of these two quantities as a
function of the time. In the early stage of the dynamics, a variety
of words are invented by agents. Because many nodes have no words in
this period, $n(t)$ keeps increasing and at a time scales as $N$, it
reaches a maximum value. We denote this maximum by $M$, and the time
$n(t)$ reaches it by $t_{max}$. After $t_{max}$, dominant word
clusters begin to engulf puny ones, and at the same time aggrandize
their sizes. This can be seen from the power-law decrease section of
$nd(t)$ and the plateau region of $n(t)$ in Fig. 2. After this
period of elimination, the system reaches a crossover point. We
denote this cross time by $t_{cross}$. After the crossover time, the
competition among the surviving words lasts until a final winner
takes up the whole network and declares the convergence of the
system. In this paper, we focus particularly on the period of word
invention($t \sim t_{max}$) and the period of word
elimination($t_{max}<t<t_{cross}$).

The first dynamical quantity we concerned is the maximum memory $M$,
which measures the minimum storage ability each agent should have,
or from another point of view, the minimum effort each agent should
make in order to reach the consensus state. Previous works
\cite{Lattice, Smallworld} had found that the maximum memory
strongly depends on the network topologies. In the following, we
provide a quantitative relationship between clustering coefficient
and the maximum memory on the basis of mean-field theory. Because
the nodes' degrees of current networks are identical, the obtained
relationship can avoid any influence of varying degrees on the
concerned quantity.
\begin{figure}
\centerline{\includegraphics*[width=0.5\textwidth]{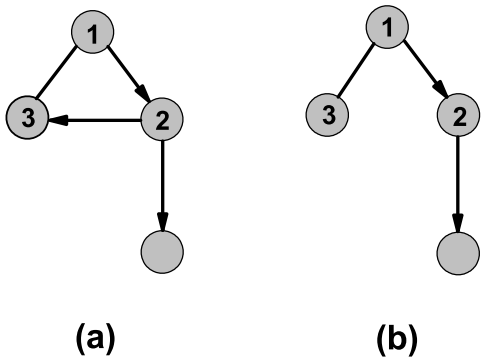}}
    \caption{The sketch map of the effect of clustering
    coefficients on the spreading process of words.}
    \label{illustrition.fig}
\end{figure}

Suppose that node $i$ is selected to be the speaker at one time, we
use $\varepsilon_i$ to denote the vocabulary size of node $i$ and
$\eta_{ij}$ to denote the size of the intersection set between the
vocabularies of the speaker $i$ and one of his neighbors $j$. If $j$
is selected to be the hearer, the probability of a successful
negotiation is $\eta_{ij}/\varepsilon_i$. Having considered that
each neighbor of $i$ has the same chance $1/k_{i}$ (here $k_{i}$ is
the degree of $i$) to be the hearer, we can formulate the successful
probability at this time:
\begin{equation}
S_{i}=\frac{1}{k_{i}\varepsilon_{i}}\sum_{j\in\Omega_j}\eta_{ij}
\end{equation}
where the sum runs over all the neighboring nodes of $i$ (this set
is indicated by $\Omega_j$). At the early stage when words begin to
spread from original nodes to the neighbors, the local success rate
of node $i$ depends strongly on its local clustering coefficient
$C_{i}$. Suppose that a node has a highly connected neighborhood,
the words it sent to its neighbors are more likely to be restricted
in its neighborhood, thus contribute to the coordinate of its
intersection sets. Figure 3 illustrates this presentation in a
simple form: node 1 sends a word to its neighbor node 2, because
hearers are randomly chosen from the speakers' neighborhood, in the
situation of figure (a), node $2$ has a chance to send this word to
node $3$, which is also a neighbor of node $1$; while in the
situation of figure (b), since the lack of triangular loops, the
word sent to $2$ can not be sent back to the neighborhood of $1$. It
is easy to generalize this simplest situation to more complex ones:
the larger the clustering coefficient of a node, the larger the
intersection set between its vocabulary and vocabularies of its
neighbors. According to Eq. (1), larger intersection set,
\emph{i.e.}, higher values of $\eta_{ij}$, will result in higher
local success rate.
\begin{figure}
\centerline{\includegraphics*[width=0.5\textwidth]{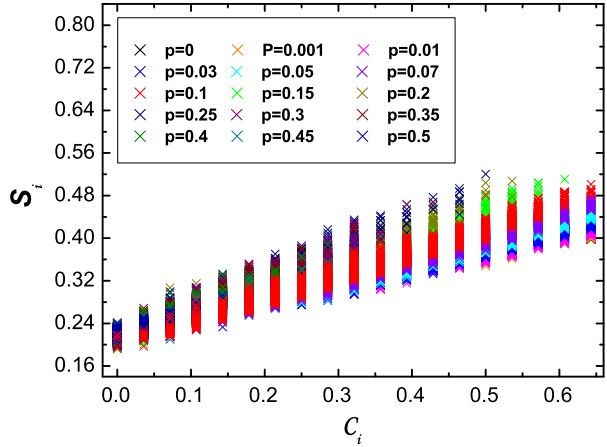}}
    \caption{(Color Online) Relationship between $S_{i}$ and $C_{i}$
    at the specific time $t_{max}$. Networks size $N=2000$. For each $p$, we generate 20 networks
    and on each network, we perform 5000 independent simulations.
    In a single simulation, once $t$ reaches $t_{max}$, the local
    success rates of nodes are calculated according to Eq.
    (1), and $S_{i}$ is obtained through averaging them over these simulations.
    Each scatter in the figure represents a node in the network sets.
    A linear positive correlation between $S_{i}$ and $C_{i}$ is suggested.}
    \label{s(i)(c(i)).fig}
\end{figure}

Figure 4 shows the numerical results of the relationship between
$S_{i}$ and $C_{i}$ at $t_{max}$ in the network ensembles of
different $p$: a linear positive correlation is obtained. We can
simply present this relationship as $S_{i}=aC_{i}+b$ by averaging
$S_{i}$ according each local clustering coefficient $C_{i}$ in the
vast ensembles of networks and get $a\approx0.35$, $b\approx0.20$
(Note that: $a$ and $b$ are implicitly involving the factor of
nodes' degree $K$ and we fix $K=8$ in the present work). Since
speakers are randomly chosen from the network, the success rate of
the step $t_{max}$ can be written as
$S(t_{max})=\frac{1}{N}\sum_{i}S_{i}$, thus we can write
$S(t_{max})=0.35C+0.20$ in the network ensembles with different
global clustering coefficients.

To get the relation between clustering coefficient and the maximum
memory, we can first expect that the total number of words $n(t)$
and the success rate $S(t)$ satisfies the following rate equation:
\begin{equation}
\frac{d n(t)}{dt}=-2S(t)(\frac{n(t)}{N}-1)+1-S(t) ,
\end{equation}
where the first item in the right side denotes the reduction of
total words with the success rate $S(t)$ and the second item denotes
increasing one word with unsuccessful rate $1-S(t)$. Here, we adopt
the mean field approximation that each node has the same number of
words $n(t)/N$ (Note that: in the highly heterogeneous networks, the
mean-field approximation may be invalid). Thus, each node will lose
$n(t)/N-1$ words and keep the successful one when the negotiation is
successful as the dynamical rules. Furthermore, considering:
\begin{equation}
\frac{d n(t)}{dt}\mid_{t_{max}}=0 ,
\end{equation}
we obtain:
\begin{equation}
M=n(t_{max})=\frac{N}{2} \left(1+\frac{1}{S(t_{max})}\right) .
\end{equation}
Hence, the maximum memory can be obtained as:
\begin{equation}
M=\frac{N}{2}\left(1+\frac{1}{aC+b}\right) .
\end{equation}
\begin{figure}
\centerline{\includegraphics*[width=0.5\textwidth]{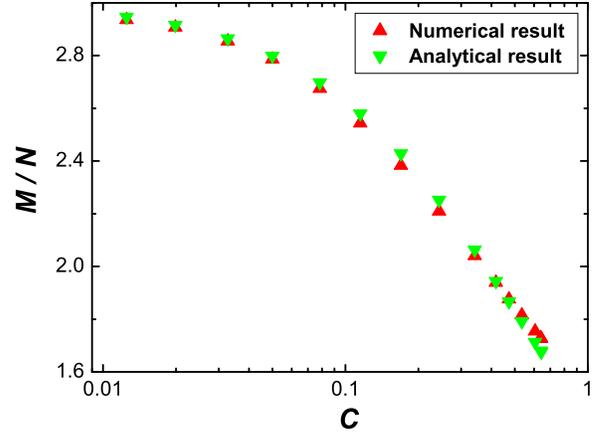}}
    \caption{(Color Online) Numerical(red) and analytical(green) results about maximum memory per agent $M/N$ as a
    function of global clustering
    coefficient $C$. $N=2000$, and $p$ is regulated to special values to obtain
    suitable clustering coefficient. On each network, 5000 independent simulations are
    performed to obtain the averaged value of $M$.}
    \label{M(P).fig}
\end{figure}
Figure 5 shows the average minimum memory per agent $M/N$, as a
function of the global clustering coefficient $C$, and provides a
consistent comparison between Eq. (5) and numerical results.
Although the linear correlation between local success rate and local
clustering coefficient is justified only in the situation of
$N=2000$, the relationship between $M$ and $C$ obtained by Eq.(5)
can be generalized to networks with different sizes. Figure 6 shows
$M$ as a linear function of the network size $N$. According to the
simulation results, we obtain $M=\alpha N$, with $\alpha$ being a
constant in a fixed $p$. This indicates that $S(t_{max})$ is
determined by rewiring probability $p$ and independent of the
network size $N$. Note that clustering coefficient of HoSW networks
is also independent of the network size $N$, as Fig. 1 has
suggested. Thus we can generalize Eq. (5) to networks with any given
sizes. By substituting the correspondent $C$ of networks with
$p=0.01$ and $p=0.1$ into Eq. (5), we obtain the correspondent
values of $\alpha = 1.71$ and $2.06$. They are very close to the
simulation results $1.74$ and $2.02$ in Fig. 6.
\begin{figure}
\centerline{\includegraphics*[width=0.5\textwidth]{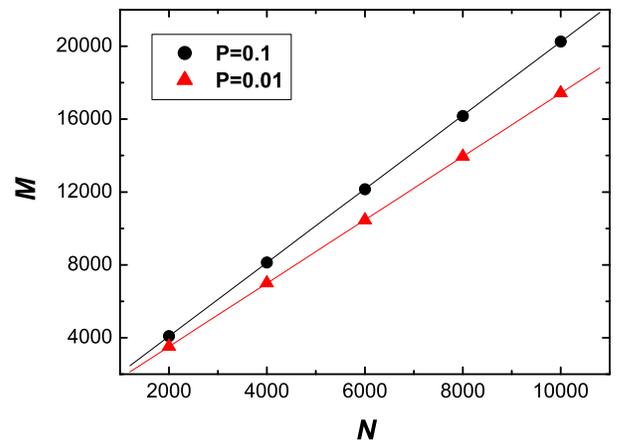}}
    \caption{(Color Online) Maximum memory $M$ as a linear function of
    network size $N$. Each value of $M$ is averaged over 5000 independent simulations.
    The functions of $p=0.01$ and $p=0.1$
    are $M=1.74N$ and $M =2.02N$ respectively.}
    \label{M(N).fig}
\end{figure}

The other dynamical quantity we concentrate on is the growing rate
of word clusters. A word cluster is a set of neighboring agents
sharing a word. Previous works had revealed that, if agents are
embedded in low-dimensional lattices ($p\ll1$), the convergence
process of the system can be analogous to a coarsening of such
clusters. The growing behavior of word clusters can be expressed as
$g(t)\sim(t/N)^{\gamma}$, with $\gamma=0.5$ in regular lattice
\cite{Lattice, Smallworld}. Here $g(t)$ is the average size of word
clusters at time step $t$. Different from former works, in our
situation, the dimension of network changes in a wide range, from
regular lattices to random networks. Thus it is very interesting to
investigate the behavior of word clusters in such networks, and the
relationships between their growing behavior and network topologies.
In the following, we also use $g(t)$, the average size of word
clusters, to describe the evolution of dynamics.
\begin{figure}
    \centerline{\includegraphics*[width=0.5\textwidth]{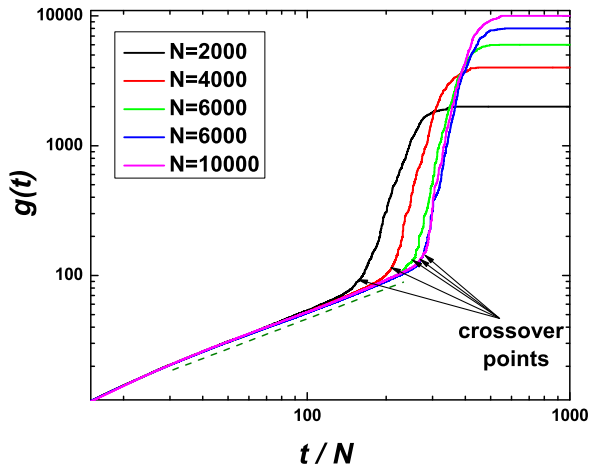}}
    \caption{(Color Online) Average cluster size $g(t)$ as a
    function of time. $p=0.1$. Each curve is separated into two distinct
    sections by a crossover time. In the first section, a power-law increase of $g(t)$ can be observed.}
    \label{c(t)(N).fig}
\end{figure}
Figure 7 displays $g(t)$ as a function of time. The evolution of
$g(t)$ is separated into two distinct sections by a crossover time
$t_{cross}$. Having considered that dynamics after crossover time is
highly affected by finite-size effects, we just pay attention to the
first section of the evolution, where $g(t)$ displays a power-law
growing behavior, as:
\begin{equation}
\begin{array}{ll}
    g(t)\sim(t/N)^{\gamma}, \ \ \ (\frac{t}{N}<t_{cross})
\end{array}
\end{equation}
The index $\gamma$ in equation (6) indicates the growing rate of
word clusters. Figure 7 implies that when rewiring probability $p$
reaches the order of $10^{-1}$, the growth of word clusters still
displays power-law behavior, as in the case of $p\ll1$; however, the
growing rate $\gamma$ differs from the case of $p=0$. Thus it is
quite natural to expect $\gamma$ depending on the network topology.
\begin{figure}
    \centerline{\includegraphics*[width=0.5\textwidth]{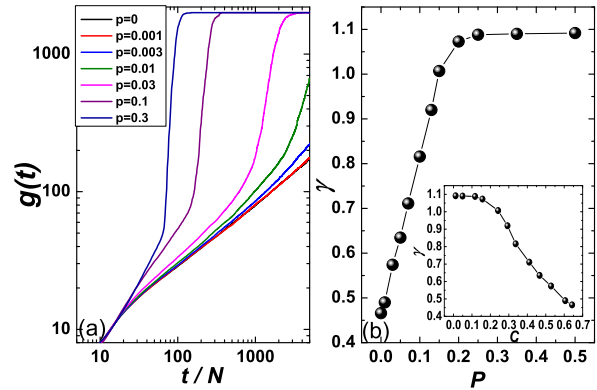}}
    \caption{(Color Online) Evolution of $g(t)$ as a function of time. Network size $N=2000$.
    Growing rates obtained from networks with different rewiring probabilities differ from each other.
    A smaller $p$ corresponds to a slower growing in the power-law section.}
    \label{c(t)(P).fig}
\end{figure}
Figure 8 shows that when rewiring probability $p$ deviates from 0,
word clusters will grow in a faster way. Note that in the model of
HoSW networks, larger $p$ yields smaller clustering coefficient, as
Fig. 1 has suggested, we can expect a negative correlation between
the growing rate and the clustering coefficient. The inset of Fig. 8
shows the relationship of these two quantities. It can be seen that
a larger clustering coefficient corresponds to a slower growing of
word clusters, and in the following paragraph, we try to give a
qualitative explanation for this phenomenon.

It is well known that the triangular loops in the network tend to
inhibit the spreading behaviors, such as the propagation of epidemic
\cite{V.M.Eguiluz}. In the context of Naming Game, a network with
many triangular loops also has a strong restriction against the
growth of word clusters. As mentioned above, clustering structures
provide a larger probability for nodes to keep words in their
neighborhoods. Consider a case of a cluster of neighboring agents
sharing a word. If they are densely connected, it is very hard for
other word clusters to invade them; thus, for a network with larger
clustering coefficient, the competition among word clusters will
take a longer time leading to a slower growing velocity of the
average cluster size.

\section{CONCLUSION AND DISCUSSION}
In this paper, we have investigated the Naming Game on homogeneous
small-world networks and focused on the influence of clustering
topology on the dynamical process. A larger clustering coefficient
allows nodes to restrict words in their neighborhoods, and thus
induces a lower maximum memory in the early stage. We obtain a
quantitative relationship between the maximum memory and the
clustering coefficient. The clustering also inhibits the growing of
word clusters, thus a network with larger clustering coefficient
requires a longer crossover time.

It is worth to emphasize that the node's degree of the network also
plays an important role in the dynamics of the Naming Game model
\cite{Non}. In this paper, we use the homogeneous networks to
eliminate the influence induced by varying the node's degree. The
relationships between node degree and current dynamics is one of the
most interesting issues that deserves further efforts to be studied.

Furthermore, recent works on the issue of Naming Game has revealed
that Naming Game model can also describe the spreading of opinions
or, from a more general viewpoint, the evolution of communication
systems. \cite{New, P.Avesani}. Although there exists some different
details when applying Naming Game to different fields
\cite{Strategy, RandomGN}, all of them are needed to capture the
essential property of this kind of dynamics, \emph{i.e.}, the system
should evolve to a consensus state without any external or global
coordination.

\begin{acknowledgments}
We acknowledge the support by the National Basic Research Program of
China (973 Program No.2006CB705500), by the National Natural Science
Foundation of China (Grant Nos. 10472116, 10532060, and 10635040),
by the Special Research Funds for Theoretical Physics Frontier
Problems (NSFC Grant Nos. 10547004 and A0524701), and by the support
from the specialized program under President Funding of Chinese
Academy of Science.
\end{acknowledgments}

\end{document}